\documentclass[aps,prl,twocolumn,showpacs,groupedaddress,superscriptaddress]{revtex4}
\usepackage{graphicx,amssymb}

\def\rmDj {\rlap{\kern 0.05em\raise 0.76ex\hbox
  {\vrule height 0.10ex width 0.28em}}D}

\begin{document}

\title{Correlated vortex pinning in Si-nanoparticle doped MgB$_{2}$}
\author{Ivica Ku\v{s}evi\'{c}}
\author{Emil Babi\'{c}}
\author{Ozren Husnjak}
\affiliation{Department of Physics, Faculty of Science, University of Zagreb, HR-10000 Zagreb, Croatia}
\author{Saeid Soltanian}
\author{Xiaolin Wang}
\author{Shi Xue Dou}
\affiliation{Institute for Superconducting and Electronic Materials, University of Wollongong, NSW 2522, Australia}

\date{\today}

\begin{abstract}
The magnetoresistivity and critical current density of well characterized
Si-nanoparticle doped and undoped Cu-sheathed MgB$_{2}$ tapes have been measured
at temperatures $T\geq 28$ K in magnetic fields $B\leq 0.9$ T. The irreversibility
line $B_{irr}(T)$ for doped tape shows a stepwise variation with a kink
around 0.3 T. Such $B_{irr}(T)$ variation is typical for high-temperature
superconductors with columnar defects (a kink occurs near the matching field $%
B_{\phi }$) and is very different from a smooth $B_{irr}(T)$ variation in
undoped MgB$_{2}$ samples. The microstructure studies of nanoparticle doped
MgB$_{2}$ samples show uniformly dispersed nanoprecipitates, which probably
act as a correlated disorder. The observed difference between the field variations of
the critical current density and pinning force density of the doped and
undoped tape supports the above findings.
\end{abstract}

\maketitle

\section{Introduction}

The discovery of superconductivity in MgB$_{2}$ compound \cite{nagamatsu}
has aroused a great deal of interest in the scientific community \cite{buzea}. Compared to
high-temperature superconductors (HTS), MgB$_{2}$ has a lower transition
temperature $T_{c}\simeq 39$ K, but its simple composition, abundance of
constituents and the absence of weak intergranular links \cite
{larbalestier,glowacki,kusevic1} make the MgB$_{2}$ a promising material for 
applications at $T\geq 20$ K, which is above $T_{c}$s of conventional
superconductors (LTS). Indeed, the simple preparation and rather high critical
currents $J_{c}$ of composite MgB$_{2}$ tapes and wires \cite
{glowacki,grasso,jin,soltanian,suo} lend strong support to these
expectations. Unfortunately, compared to practical LTS (NbTi, Nb$_{3}$Sn),
MgB$_{2}$ exhibits weak flux-pinning \cite{buzea,finnemore}, which results in
strong field dependence of $J_{c}$ and a low irreversibility field 
$B_{irr}(4.2 $ K$)\approx 8$ T \cite{buzea}.

Several techniques, such as alloying \cite{eom,feng,cimberle}, particle
irradiation \cite{bugoslavsky,babic1,eisterer,okayasu} and mechanical
processing \cite{suo,gumbel} have been employed in order to improve the flux-pinning in MgB$_{2}$, but with limited success. In particular, proton
irradiation \cite{bugoslavsky} increased $B_{irr}$ at 20 K, but also
suppressed low-field $J_{c}$, whereas alloying seems to enhance $J_{c}$, but
has little effect on $B_{irr}$ \cite{feng,cimberle}. Better results were
recently obtained by adding nanoparticles to MgB$_{2}$ \cite
{jwang,dou,xwang}. It appears that a variety of nanoparticles considerably
enhance the flux-pinning in MgB$_{2}$ over a wide temperature range 
$T\leq 30$ K. In particular, the addition of 10 wt\% of SiC nanoparticles 
 \cite{dou} yielded $B_{irr}(4.2$ K$)\gtrsim 12$ T, which is higher than that
of optimized NbTi \cite{meingast}. The actual mechanism 
of the flux-pinning enhancement upon nanoparticle doping of MgB$_{2}$ is not well understood at present.

Here we present the results for magnetoresistance $R(T,B)$ and critical
current $I_{c}(T,B)$ of MgB$_{2}$ tape doped with Si-nanoparticles, which
reveal the flux-pinning mechanism associated with nanoparticle doping. In
particular, $B_{irr}(T)$ of doped sample shows a kink at $B_{irr}\approx 0.3$
T, which is the signature of vortex pinning at correlated defects \cite
{blatter}, whereas no kink is observed in undoped sample. The variation of
critical current and pinning force density $F_{p}=J_{c}B$ with the field and
temperature also show different pinning mechanisms in doped and undoped MgB$%
_{2}$, respectively.

\section{Experimental procedures}

Cu-sheathed MgB$_{2}$ tapes were prepared by in-situ powder-in-tube method 
 \cite{soltanian}. In the doped tape, in addition to Mg and B, 5 wt\% of
Si-nanoparticles with an average size $\sim 50$ nm was added. A low sintering
temperature (670--690$^{\circ}$ C) and a short sintering time (several minutes) were employed \cite{xwang}
in order to avoid diffusion of Cu into the MgB$_{2}$ core \cite{majoros}.
This resulted in rather porous, low density ($\sim 50\%$) cores. The core
cross-sections were eliptical with areas $4.95\cdot 10^{-3}$ and $4.8\cdot
10^{-3}$ cm$^{2}$ for the doped and undoped tape, respectively. The sample
lengths were approximately 1.5 cm and the voltage and current leads were
soldered on Cu-sheathing. The magnetoresistance was measured with
low-frequency ac method \cite{kusevic1,babic1} for $T\geq 28$ K in magnetic
field $B\leq 0.9$ T perpendicular to a broad face of the tape and the
current direction. $I_{RMS}=1$ mA was used and the voltage resolution was
0.3 nV. Critical currents were measured on samples used in $R(T,B)$
measurements with the pulse method (saw-tooth pulse with duration less than
10 ms and peak current of 200 A \cite{soltanian}).

\section{Results and discussion}

The variation of the resistance with temperature ($28\leq T\leq 300$ K) for our
undoped and Si-doped tape (Fig. 1) are typical for Cu-clad MgB$_{2}$ wires
 \cite{majoros}, with a larger resistance of the doped sample due to a larger
distance between its voltage contacts.

\begin{figure}[tbp]
\includegraphics[width=8.6cm]{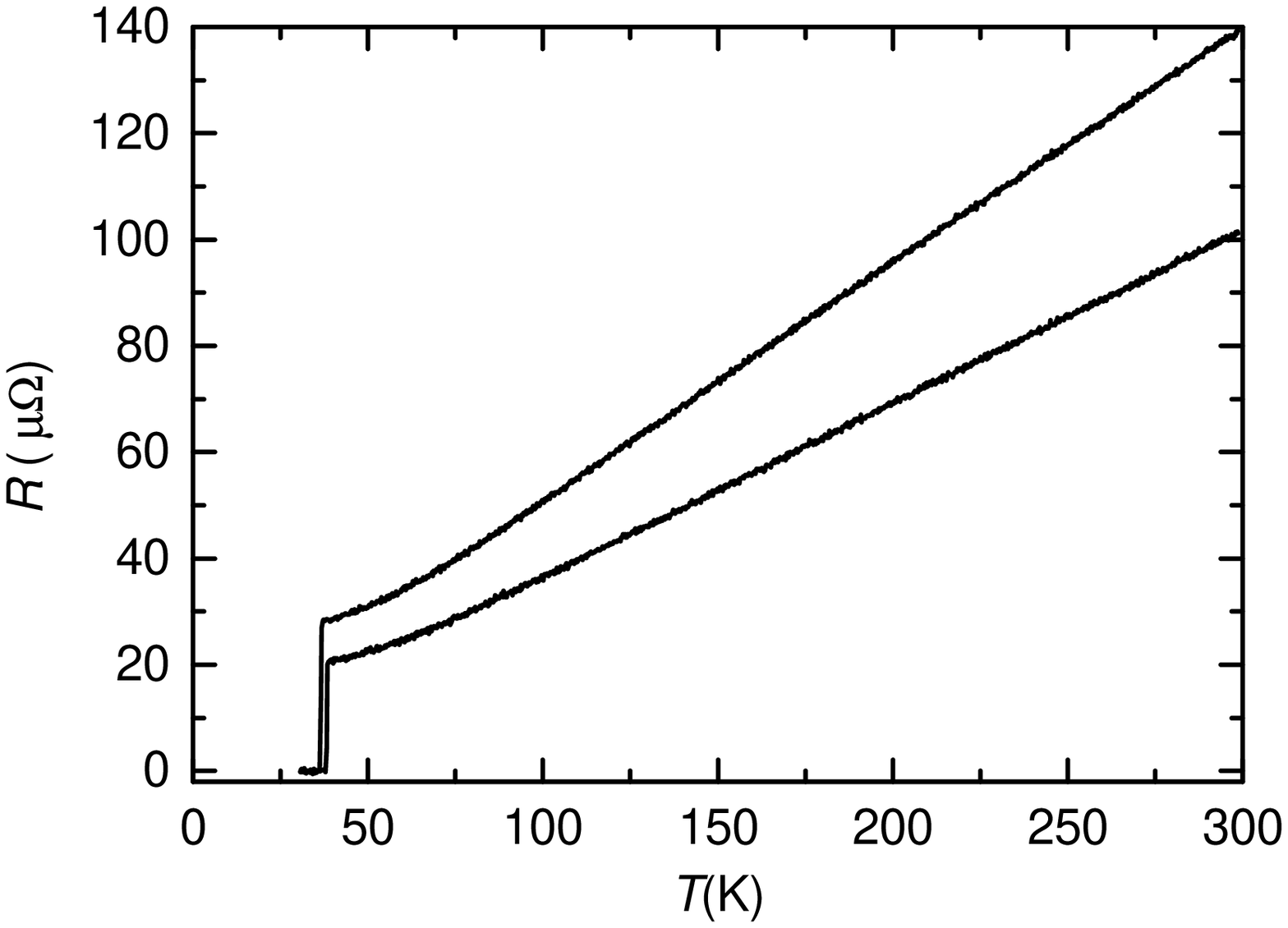} 
\begin{caption}{Temperature variation of the electrical resistance for the 
undoped (lower) and doped (upper curve) sample.}
\end{caption}
\end{figure}

\begin{figure}[tbp]
\includegraphics[width=8.6cm]{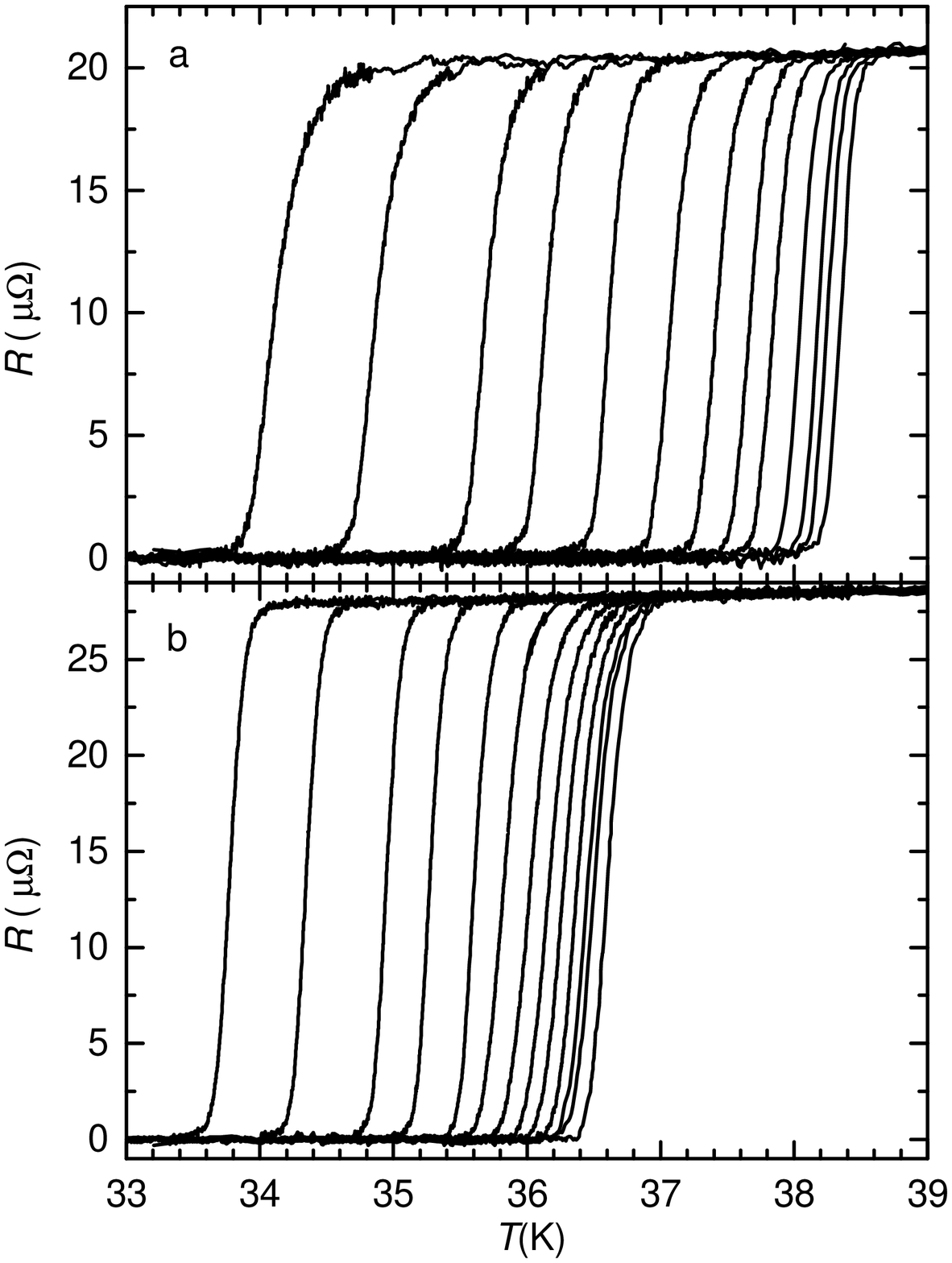} 
\begin{caption}{Temperature variation of the electrical resistance in 
magnetic fields $B=0$, 0.01, 0.02, 0.04, 0.07, 0.1, 0.14, 0.2, 0.3, 0.4, 
0.5, 0.6, 0.7, 0.8 and 0.9 T for the a) undoped and b) doped sample.}
\end{caption}
\end{figure}

Fig. 2 compares the superconducting transitions in fields $B\leq 0.9$ T for
undoped and doped sample. As in other composite superconductors \cite{babic2},
the shape of these transitions is affected by Cu-sheathing. However, the
onset of resistance (hence $T_{c}(R\rightarrow 0)=T_{c0}$) is not affected
by sheathing \cite{babic2}, and the zero-field $T_{c0}=38.2$ K\ for the
undoped tape (Fig. 2a) is typical for bulk MgB$_{2}$ samples \cite
{buzea,kusevic1,majoros}. A strong shift of its $T_{c0}$ with magnetic field
(i.e. $T_{irr}(B)$) reflects a weak flux-pinning in the undoped MgB$_{2}$.
For the doped sample (Fig. 2b), zero-field $T_{c0}=36.4$ K is lower than
that of the undoped one, but the shift of its $T_{c0}$ with field is considerably 
smaller, which indicates an enhancement of flux-pinning (the expansion of
the vortex-solid regime). Furthermore, values of $T_{c0}$ for the doped
sample in $B\lesssim 0.3$ T are compressed within a rather narrow
temperature interval, whereas those for the undoped one are more evenly
spread throughout the explored field range.

Fig. 3 compares the irreversibility fields $B_{irr}(T)$ (defined by using
the low-resistivity criterion $\rho _{c}=5$ n$\Omega $cm) for our samples.
For the undoped tape, both the magnitude and temperature variation of $%
B_{irr}$ are the same as the literature data for MgB$_{2}$ samples \cite
{buzea,kusevic1,eisterer,majoros,gioacchino}. In particular, our values of $%
B_{irr}(T)$ are equal to those obtained from the onset of the third harmonic
in the low-frequency ac susceptibility of a dense MgB$_{2}$ sample \cite
{gioacchino}. Approximately linear, $B_{irr}(T)$ variation for $T\leq 36$ K
extrapolates to $B_{irr}(4.2$ K$)\approx 8.4$ T, which is a typical value
for bulk MgB$_{2}$ \cite{buzea}.

\begin{figure}[tbp]
\includegraphics[width=8.6cm]{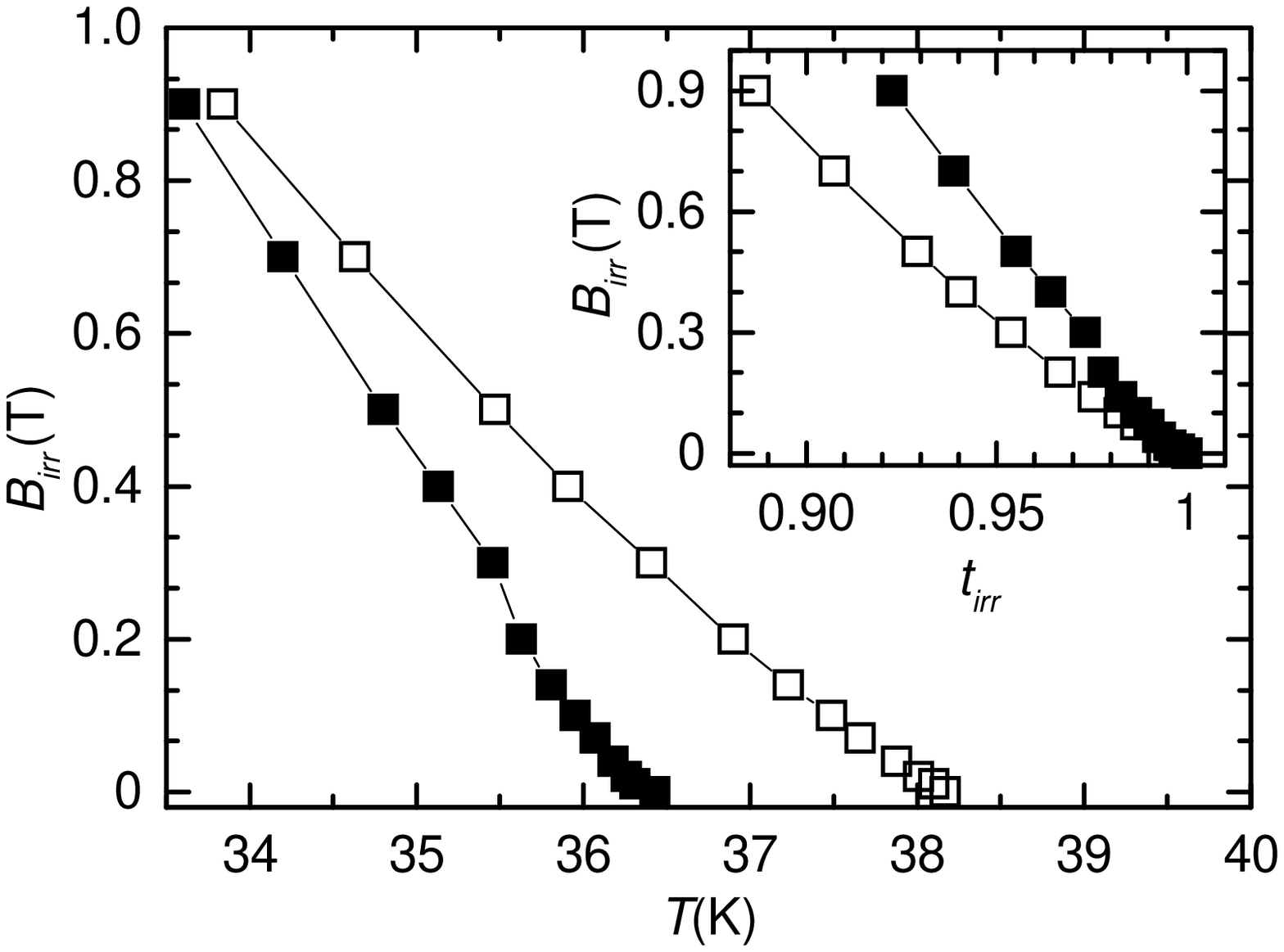} 
\begin{caption}{Temperature dependence of the irreversibility field $B_{irr}$ 
for the undoped (empty) and doped (full squares) sample. Inset: the same 
dependence, but vs. the reduced temperature $t_{irr}=T_{irr}(B)/T_{irr}(0)$.}
\end{caption}
\end{figure}

The $B_{irr}(T)$ variation for the doped tape is very different from that of
the undoped one (Fig. 3). Here, $B_{irr}$ increases rapidly with decreasing
temperature down to 35.5 K, and shows slower, linear variation for $T\leq 35$
K. Such a stepwise $B_{irr}(T)$ variation is specific for HTS containing
columnar defects \cite{blatter,krusin,beek,daignere}, where the crossover in $%
B_{irr}(T)$ occurs around the matching field $B_{\phi }$, which is the field
at which the vortex and columnar defect density $n_{\phi }$ are equal ($%
B_{\phi }=n_{\phi }\Phi _{0}$, $\Phi _{0}$ being the flux quantum \cite
{beek,daignere}). This crossover occurs because the pinning of interstitial
vortices for $B_{irr}>B_{\phi }$ is less weaker than for vortices
residing onto the columns for $B_{irr}<B_{\phi }$.

From our crossover field $B_{c}\approx B_{irr}(35.5$ K$)\simeq 0.3$ T we
estimate $n_{\phi }\approx 1.4\cdot 10^{14}$ m$^{-2}$, and the average
distance between defects $\sim 80$ nm. The microstructural studies of the
nanoparticle doped MgB$_{2}$ \cite{jwang,dou,xwang} show finely dispersed
precipitates within the MgB$_{2}$ matrix with sizes $\sim 10$ nm. For Si and
SiC doped MgB$_{2}$ \cite{dou,xwang} these precipitates are mainly Mg$_{2}$Si
phase, and their average spacing is comparable to that estimated above.
Therefore in our tape Mg$_{2}$Si nanoprecipitates, resulting from the
reaction of Si-nanoparticles and Mg during the sintering, act analogously to
columnar defects in HTS. This outcome appears rather surprising considering
different nature and geometries of precipitates and columns, as well as the
different nature of vortices \cite{blatter,eskildsen} in these materials.
However, the matching effects are common in type-II superconductors \cite
{petermann} and are not specific only to HTS.

A linear variation of $B_{irr}(T)$ for $T\leq 35$ K in the doped sample
extrapolates to $B_{irr}(4.2$ K$)\simeq 11.5$ T, which is consistent with
the other results for nanoparticle-doped MgB$_{2}$ \cite{jwang,dou,xwang},
and is higher than $B_{irr}(4.2$ K) for NbTi. However, for $T>33$ K, $B_{irr}
$ of the doped sample is lower than that for the undoped sample, which is
entirely due to its lower zero-field $T_{c0}$. Indeed, a plot of $B_{irr}$
vs. reduced temperature $t_{irr}=T_{irr}(B)/T_{irr}(0)$ for both samples
(inset to Fig. 3) shows that for all values of $t_{irr}$, $B_{irr}$ of the
doped sample is higher than that of the undoped one. Therefore, vortex
pinning in nanoparticle doped MgB$_{2}$ is enhanced with respect to that in
undoped MgB$_{2}$ at all reduced temperatures.

Different vortex pinning mechanisms in our tapes imply also different field
variations of their $J_{c}$ and $F_{p}=J_{c}B$. Fig. 4 compares the $J_{c}(B)
$ variations of our samples for $T\geq 33$ K. The undoped tape (Fig. 4a)
shows approximately exponential $J_{c}(B)$ variation, which is typical for
MgB$_{2}$ samples \cite{buzea,finnemore,bugoslavsky,dou}. At low temperatures
(high $I_{c}$), large self-field $\mu _{0}H_{s}$ ($H_{s}\simeq I_{c}/c$, 
where $c$ is the circumference of the core) makes $J_{c}(B<\mu _{0}H_{s}$) nearly
constant, whereas at elevated fields ($B\rightarrow B_{irr}$) $J_{c}$
rapidly decreases to zero. From the experimental $J_{c}(B,T)$ curves (Fig.
4a) we obtained $F_{p}(B,T)$ ones, from which we determined the fields $%
B_{\max }(T)$ at which the volume pinning force density reaches its maximum
value $F_{p\max }=J_{c}B_{\max }$. The field $B_{\max }$ is an important
parameter of the vortex pinning within the vortex-solid phase. In
particular, in the case of dominant vortex pinning mechanism, the ratio 
$B_{\max }/B_{irr}$ may reveal this mechanism. For the undoped tape we found $%
B_{\max }/B_{irr}\approx 0.21$, which is similar to that observed in Nb$_{3}$%
Sn \cite{hughes} and is consistent with a commonly accepted grain boundary
pinning mechanism for a bulk MgB$_{2}$ \cite{larbalestier}. In spite of a
probably common vortex pinning mechanism in both bulk MgB$_{2}$ and Nb$_{3}$%
Sn, vortex pinning in MgB$_{2}$ is apparently weaker (lower $B_{\max }$ and $%
B_{irr}$) than that in Nb$_{3}$Sn. The probable reason for that are larger
grains, clean and narrow grain boundaries, and quite a large coherence length
 \cite{eskildsen} in MgB$_{2}$.

\begin{figure}[tbp]
\includegraphics[width=8.6cm]{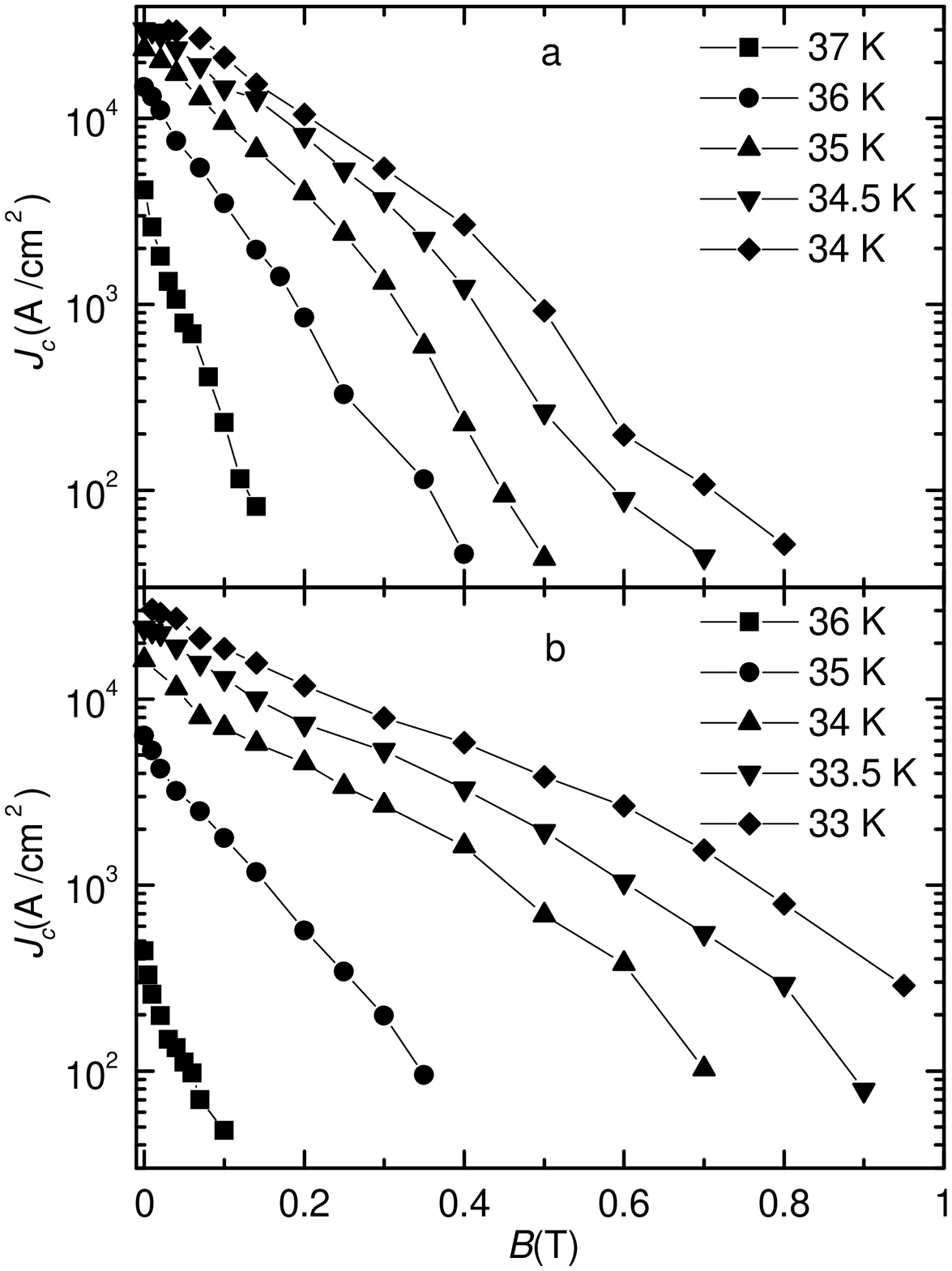} 
\begin{caption}{Dependence of the critical current density $J_{c}$ on magnetic field $B$ at denoted temperatures for the a) undoped and b) doped sample.}
\end{caption}
\end{figure}

The $J_{c}(B)$ variation of nanoparticle doped MgB$_{2}$ sample (Fig. 4b) is
very different from that for the undoped tape (Fig. 4a). The S-shaped $%
J_{c}(B)$ curves of the doped tape are reminiscent of those observed in
HTS films, tapes and crystals containing columnar defects \cite{blatter}.
Further, for the same reduced temperature $t=T/T_{c}$, the decrease of $J_{c}
$ with $B$ in the doped tape is considerably smaller than that for the
undoped one. Accordingly, the fields $B_{\max }(t)$ are enhanced with respect
to those of the undoped tape, which shows that nanoparticle doping enhances
vortex pinning throughout the vortex-solid regime \cite{blatter}.
Furthermore, the enhancement of $B_{\max }(t)$ in the doped tape is larger
than that of $B_{irr}(t)$, which results in $B_{\max }(t)/B_{irr}(t)\approx
0.29$ for the doped tape. Such $B_{\max }/B_{irr}$ ratio is unlikely to
arise only from the grain boundary pinning \cite{hughes} and was earlier
observed for HTS tapes \cite{babic3} with a modest density of columnar
defects ($B_{\phi }\lesssim 0.2$ T). Therefore, we propose that $B_{\max
}/B_{irr}\approx 0.29$ arises from the competition of two pinning mechanisms
(for example, a grain boundary pinning and a core pinning at
nanoprecipitates) as was the case in HTS tapes. A detailed investigation of $%
J_{c}(B)$ curves for a number of temperatures extending over a broad
temperature range (which requires $I>200$ A) is necessary in order to solve
this problem.

In spite of 50\% porosity, our tapes have large self-field $J_{c}$s (Fig.
4), which increase rapidly with decreasing temperature ($J_{c}(t)\simeq
J_{c}(0)(1-t)^{n}$, with $n\approx 1.5$). In particular, the observed $%
J_{c}(0.9T_{c})\approx 40$ kA/cm$^{2}$ for both tapes extrapolate to $%
J_{c}(20$ K$)\approx 350$ kA/cm$^{2}$, the value which was confirmed by the
magnetic measurements of $J_{c}(20$ K) \cite{xwang}. Therefore, fully dense
MgB$_{2}$ tapes are expected \cite{kusevic1,jwang} to reach $J_{c}(20$ K$%
)\sim 10^{6}$ A/cm$^{2}$, which is above $J_{c}(4.2$ K) for the best
Bi2223/Ag tapes.

In summary, we have shown that a uniform dispersion of Mg$_{2}$Si
nanoprecipitates (resulting from the addition of Si-nanoparticles to Mg and
B powders \cite{dou,xwang}) not only enhances the flux-pinning in MgB$_{2}$
samples, but also introduces an additional pinning mechanism. In particular,
we observed a step-wise variation of $B_{irr}(T)$ in nano-Si doped MgB$_{2}$
tape with a kink around $B_{\phi }\simeq 0.3$ T, which is reminiscent of the
vortex pinning at correlated disorder in HTS \cite
{blatter,krusin,beek,daignere}. We also observed a corresponding difference
in the shapes of $J_{c}(B)$ and $F_{p}(B)$ curves for the doped and undoped
tape respectively. Although our results were obtained for MgB$_{2}$ tape
doped with Si nanoparticles only, we believe that the above conclusions hold also for other MgB$_{2}$ samples doped with different types of nanoparticles \cite{jwang,dou,xwang}, providing that these nanoparticles form uniformly
dispersed non-superconducting nanoprecipitates.

\begin{acknowledgments}
We thank Ms Mino Delfany for the praparation of samples.
\end{acknowledgments}

\bibliographystyle{apsrev}
\bibliography{correl}

\end{document}